\documentclass{elsart}
\journal{Astroparticle Physics}

\usepackage{amssymb}
\usepackage{amsmath,amsfonts,verbatim,graphicx,float}
\usepackage{subfigure}
\usepackage{textcomp}
\usepackage{cite}

\newfont{\tensy}{cmsy10}

\begin{document}
 
\begin{frontmatter}

\title{A method to measure the mirror reflectivity 
of a prime focus telescope}

\author[]{R. Mirzoyan\corauthref{cor1}},
\corauth[cor1]{Corresponding author.}
\ead{razmik@mppmu.mpg.de}
\author[]{M. Garczarczyk},
\author[]{J. Hose},
\author[]{D. Paneque}

\address{Max-Planck-Institut f\"ur Physik, F\"ohringer Ring 6, 80805 
Munich, Germany}

\begin{abstract}
We have developed a method to measure the mirror reflectivity of telescopes. While it is relatively easy to measure the local reflectivity of the mirror material, it is not so straightforward to measure the amount of light that it focuses in a spot of a given diameter. Our method is based on the use of a CCD camera that is fixed on the mirror dish structure and observes simultaneously part of the telescope's focal plane and the sky region around its optical axis. A white diffuse reflecting disk of known reflectivity is fixed in the telescopes focal plane. During a typical reflectivity measurement the telescope is directed to a selected star. The CCD camera can see two images of the selected star, one directly and another one as a spot focused by the mirror on the white disk.
The ratio of the reflected starlight integrated by the CCD from the white disk to the directly measured one provides a precise result of the product of (mirror area $\times$ mirror reflectivity).
\end{abstract}

\begin{keyword}
mirror reflectivity \sep telescopes \sep IACT.
\PACS 42.15.Eq \sep 42.15.-i 
\end{keyword}
\end{frontmatter}

\section{Introduction}

Very high-energy (VHE) gamma ray astronomy is making fast progress. Since the first detection of a VHE gamma signal from the Crab Nebula by the Whipple collaboration in 1989~\cite{Weekes} almost 40 sources have been discovered. The new generation of Imaging Air Cherenkov Telescopes (IACT) that are currently operating or are just beginning to start their operation, hold the promise to further increase the number of sources in the next few years. \\

The reflectivity of the mirror is one of the important parameters in the performance of an IACT. The mirror reflectivity has a direct impact on the absolute scale of measured energy and the precision of measured flux of gamma rays from sources. In addition, the mirror reflectivity is important in the determination of the energy threshold of an IACT. The latter is inversely proportional to the quantity (mirror area $\times$ mirror reflectivity)~\cite{Mirzoyan}. It is desirable to constantly monitor the reflectivity of the mirror of a telescope because of the aging and because of the dust and other possible deposits on the mirror surface. We have developed an experimental method that allows one to measure the mirror reflectivity with a precision of $\leq 2\%$ in a few minutes time. Below we report on the details of this method and the results from the measurements on the MAGIC telescope~\cite{magic}.

\section{Description of the method} 
\label{chapter:method}

\begin{figure}[htb]
\begin{center}
\includegraphics[width=0.45\linewidth]{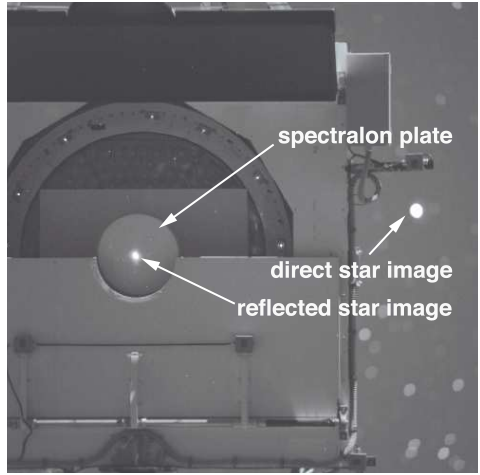}
\caption{Image taken with the SBIG CCD camera mounted on the reflector of the MAGIC telescope. The diffuse reflecting spectralon disk is fixed on the lower lid of the imaging camera. The direct image of a tracked star as well as its reflected spot on the disk can be seen in the FOV of the CCD.}
\label{fig:fov}
\end{center}
\end{figure}

A CCD camera of high resolution and sensitivity is fixed near the optical axis of the reflector, at a place where it is just out of the shadow cast by the imaging camera for on-axis beam of light. The CCD camera shall have a large enough field of view (in our case $\sim 4^\circ$ in diameter) covering part of the focal plane of the telescope and the sky region around the imaging camera, as it is shown in Fig.~\ref{fig:fov}. For the reflectivity measurement a white diffuse reflecting disk is placed in the focal plane of the telescope. The telescope is sent to a selected relatively bright star (or alternatively to a distant artificial source of light, like for example, a light emitting diode (LED)). In this setup the CCD camera can see two images of the same star: its direct image and the image reflected and focused by the mirror on the white disk. Depending on the mirror size and its focal length the amount of light from two images can be quite different. One needs a CCD camera with high dynamic range in order to resolve the direct and reflected images with high precision, in the absence of saturation effects.

The CCD camera shall be focused on the focal plane of the telescope. Because of this the direct image of the star from infinity will appear somewhat smeared out and that helps to avoid the saturation effects.

Simple calculation of the ratio of the reflected signal to the directly measured one shows that one can measure the product reflectivity times mirror surface area $R_{mirror} \times A_{mirror}$ by using the relation:
\begin{eqnarray}
R_{mirror} \times A_{mirror} = \frac{Signal_{reflected}}{Signal_{direct}} 
\cdot \frac{\pi \cdot d^{2}}{\cos{\alpha}} \cdot \frac{1}{R_{diffuser}}
\label{eq:rtimesa}
\end{eqnarray}

\noindent
where $Signal_{direct}$ and $Signal_{reflected}$ are the integrated amounts of light (measured charge) extracted from the same image frame of the CCD; $R_{diffuser}$ is the reflectivity of the white disk placed in the telescopes focal plane, $\alpha$ is the angle between the normal of the white disk and the line connecting its centre with the CCD camera lens and $d$ is the distance between the latter and the disk. The reflectivity of the chosen white disk shall be measured under laboratory conditions and one needs to check that the reflected light intensity follows the Lambert's cosine distribution. 

\section{Results} 
\label{chapter:results}

We have installed the above described reflectivity measuring system on the MAGIC telescope. It is allowing us to continuously monitor the reflectivity and the point spread function of the mirror. 

\begin{figure}[h]
\centering
\includegraphics[width=0.4\linewidth]{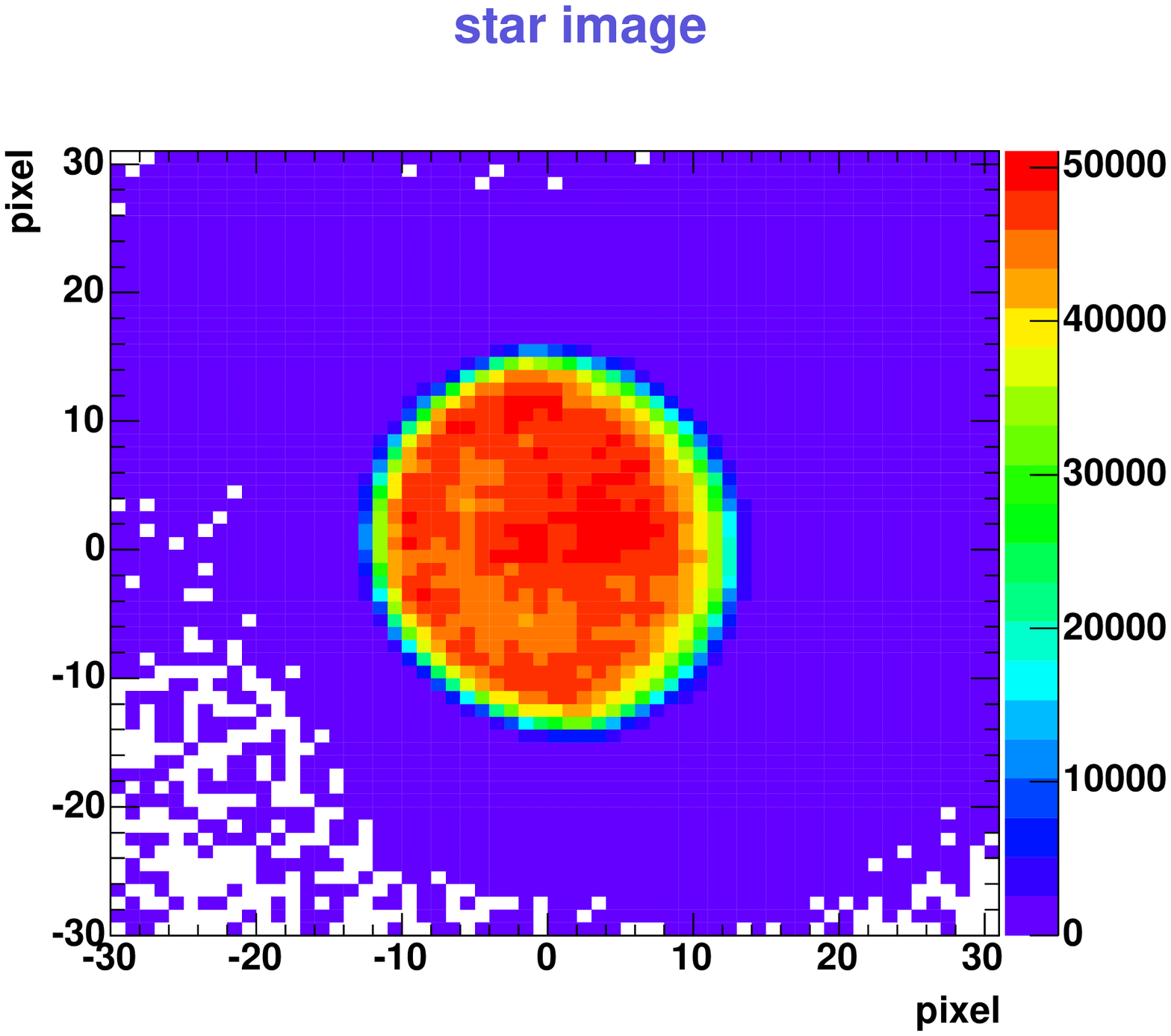}
\includegraphics[width=0.4\linewidth]{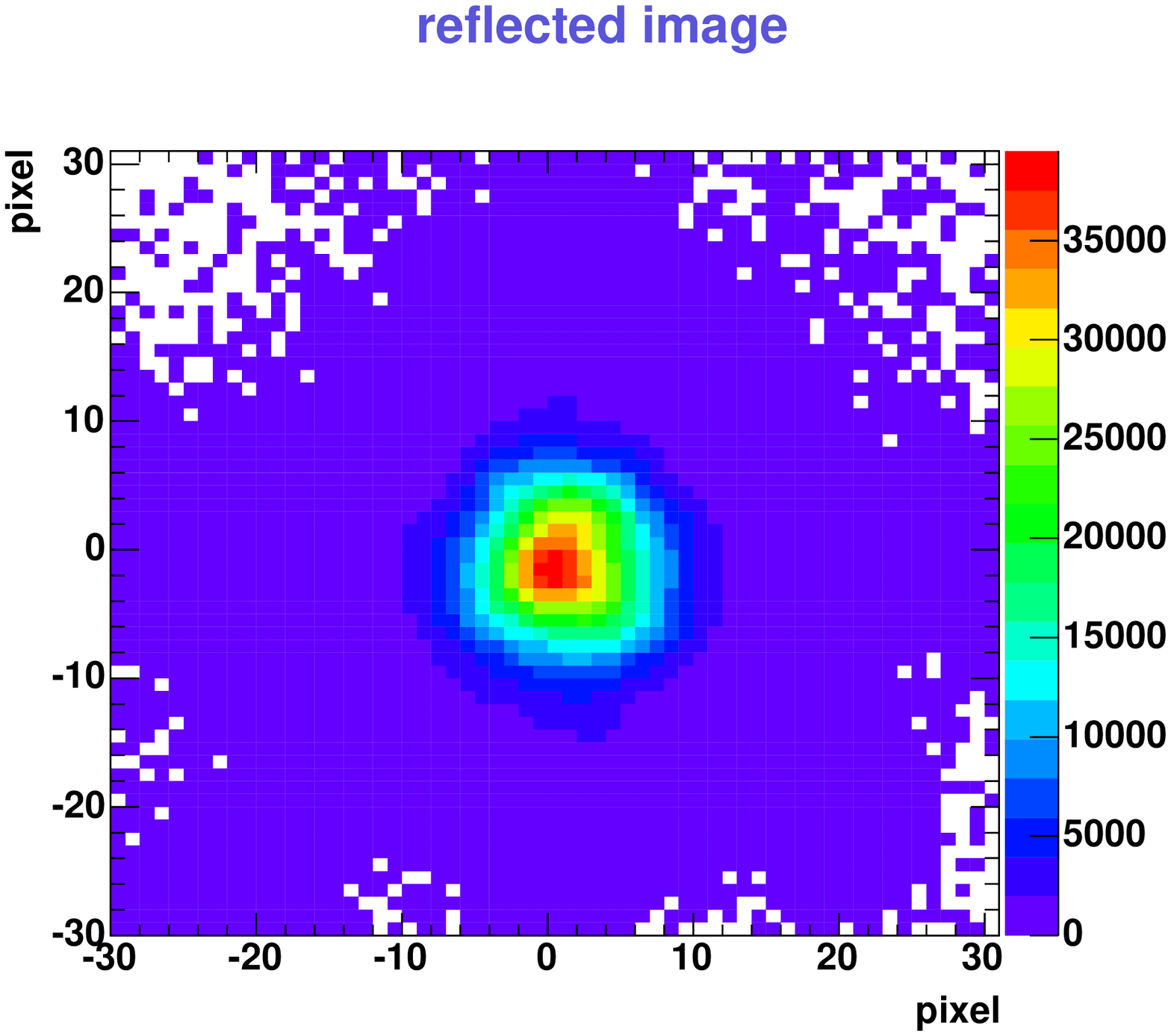}
\caption{Images of the direct star (left) and reflected spot (right) taken with the CCD camera. Note: due to the high dynamic range the star image is not saturated.}
\label{fig:measurement}
\end{figure}

For this setup we have purchased a STL-1001E astronomical CCD camera from the company SBIG. It has $16 \, \mathrm{bits}$ resolution and a matrix of $1024 \times 1024$ pixels with the size 24 \textmu m $\times$ 24  \textmu m each. We are using a Nikon MF 180/2.8 lens for focusing the CCD camera at the $17 \, \mathrm{m}$ focal plane of the telescope. In this setup $2.34 \, \mathrm{mm}$ in the focal plane are imaged onto one pixel of CCD. Star images from infinity are somewhat blurred which allows longer exposure times before saturation. The camera includes a filter reel inside a housing that is equipped with the optical V, B, G and R filters. These are allowing to perform spectral reflectivity measurements. A white $30 \, \mathrm{cm}$ diameter and $6 \, \mathrm{mm}$ thick custom made disk from spectralon~\cite{Razmik} is permanently fixed to the lower lid of the camera. It is shown in reference~\cite{Razmik} that this material has $(95-99) \%$ reflectivity for the wavelength band $(340-1000) \, \mathrm{nm}$ and at $300 \, \mathrm{nm}$ it still has, depending on the production batch, $(92-97) \%$ reflectivity. During the reflectivity measurements one remotely closes the lower lid of the camera thus setting the spectralon disk in the target position in the focal plane (the position of the disk can be adjusted along the telescopes axis for the best focus) and exposing it to starlight. For demonstration purposes we show below a reflectivity measurement performed on the ${2.23}^m$ star {\it Alphecca}. The CCD exposure time was 10 seconds. On Fig.~\ref{fig:measurement} left one can see the direct image of the star and on the right one can see the reflected image. The signals $Signal_{reflected}$ and $Signal_{direct}$ were measured by integrating the contents of the illuminated by the star pixels of the CCD after background subtraction. The quantities $\alpha = 4.69 \pm 0.03^\circ$ and $d = 17045 \pm 3 \, \mathrm{mm}$ were measured on the telescope. In the laboratory we have measured a $R_{diffuser}$ of $(96 \pm 1)\%$ for the spectralon disk. We also measured that the light reflected from the disk can be described by the Lambert's cosine low. By taking the effective mirror area of $A_{mirror} = 230 \, \mathrm{m}^2$ and allowing for 2.5\% shadowed area due to the imaging camera and its holding masts and steel cables, we obtained a value of $R_{mirror} = (80.4 \pm 2)\%$ for the effective reflectivity of the mirror of the MAGIC telescope. No filter was used in this measurement. We also measured the reflectivity in the wavelength bands $(380-520) \, \mathrm{nm}$, $(430-580) \, \mathrm{nm}$ and $(580-700) \, \mathrm{nm}$ and found out that the reflectivity was $(81.0 \pm 2)\%$, $(84.3 \pm 2)\% $ and $(81.0 \pm 2)\%$ correspondingly. 

We performed an independent measurement of $R_{mirror}$ using a $6 \, \mathrm{W}$ halogen lamp located at $\approx 1 \, \mathrm{km}$ distance from the telescope. We compared the light intensity arriving at the telescopes reflector with the one being focused onto the camera plane. The former was measured by a large $(7.84 \,\mathrm{cm}^2)$ PIN diode, whereas the latter was measured by means of a matrix of 19 small ($1\, \mathrm{cm}^2$ each) PIN diodes. The value of $R_{mirror}$ obtained with this method was about $(71 \pm 5.5)\%$. The difference of this result from the above obtained numbers could be explained by the non-identical response curves of the used diodes as well as by the fact that the light from the distant lamp was not homogeneous over the reflectors area.

\section{Discussion} 
\label{chapter:discussion}

The above described method provides a measurement of $R_{mirror} \times A_{mirror}$ with high precision. It is namely this product that has a physical meaning and is usually used in the Monte Carlo simulations. If one wants to measure the reflectivity alone, then one needs simply to measure or to calculate the exact mirror area that is exposed to light and to normalize the $R_{mirror} \times A_{mirror}$ to that number.

The reflecting disk in the focal plane can be made of any appropriate material like, for example, the above mentioned spectralon, or other materials that can be painted white for best reflectivity. It is important to measure carefully in the laboratory the reflectivity of the disk and its diffuse scattering characteristics. Measurements in regular time intervals can allow one to follow a possible degradation of the disk reflectivity because of, for example, dust deposit.

In the case of the MAGIC telescope the spectralon disk is inside the imaging camera housing and is protected from weathering. Only during the observations with the telescope it is exposed to the sky.
The entire reflectivity measurement can be done in few minutes time. For best results one needs to use a CCD camera of dynamic range $> 60 \, \mathrm{db}$. The high dynamic range allows one to accurately subtract the background illumination, to work with bright stars (or alternative light sources) of different magnitudes, providing very high signal/noise ratios. Our measurements showed that with the used camera it was even possible to work with partial moon illumination. Stars of magnitudes up to four can easily be used as targets. 

The absolute calibration of the CCD camera is not important because the method is using the ratio of two measurements in the same recorded image. High homogeneity and low charge transfer losses of the CCD chip could become important when a precision of $\leq 1\%$ is needed.

The reflectivity of a mirror depends on the wavelength of light. A colour filter wheel in the CCD camera can allow one to measure the reflectivity at different wavelengths. By selecting a target star of an appropriate temperature one can optimize the measuring time for the given wavelength band. For narrow bands one will need to integrate somewhat longer but in any case such a measurement can be done within a few minutes time. Also alternative light sources as LEDs and lasers can be used for covering a wider range of wavelengths and intensities. 

The wavelength dependent measurements of the mirror reflectivity of MAGIC have shown values within $(80.4-84.3 \pm 2)\% $ for the wavelength band $(380-700) \, \mathrm{nm}$. The knowledge of this wavelength dependency would further improve the precision of gamma ray flux measurements.

\section{Conclusions}
\label{chapter:conclusions}

We have developed a method to measure the product $R_{mirror} \times A_{mirror}$ of a telescope. This method allows one to measure the above quantity at desired wavelengths in a few minutes time with accuracy better than 2\%. By normalizing the $R_{mirror} \times A_{mirror}$ to the real mirror area that is exposed to light one obtains the $R_{mirror}$.

\end{document}